\documentclass[aps,prb,twocolumn,superscriptaddress,showpacs]{revtex4}

\usepackage{amsmath}
\usepackage{bm}
\usepackage{graphicx}

\begin{document}
\title{A terahertz molecular switch}

\author{P.\ A.\ Orellana}
\affiliation{Departamento de F\'{\i }sica, Universidad Cat\'{o}lica del Norte,
Casilla 1280, Antofagasta, Chile}
\author{F. Claro}
\affiliation{{\it Facultad de F\'{\i }sica, }Pontificia Universidad Cat\'{o}lica de Chile, Casilla 306, Santiago 22, Chile}

\begin{abstract}
We present time-dependent results describing the current through a molecular
device, modeled as a complex with two active centers connected to leads
under bias. We show that, at a properly adjusted external voltage, a passing
terahertz electromagnetic pulse may cause a transition between states of
finite and negligible current, suggesting that the system might be useful as
a nanoscopic switch in the terahertz range. A phase diagram defining the
bias region in which the transition takes place within a short time is
given. As described, the physical processes involved are of an entirely
different nature than those in ordinary photodetectors.
\end{abstract}

\pacs{73.40.Gk, 73.63.-b, 85.65.+h, 74.78.Na}

\maketitle

Individual molecules as well as molecular complexes are increasingly being
perceived as possible electronic device elements, constituting great promise
in the process of miniaturization. Some successes have already been reported
along this line\cite{dekk,joac,park,lian}. Among recent examples are
molecules that exhibit big on-off current ratios and a large negative
differential resistance in two-terminal transport\cite{chen,chen1}, behaving
much in the way some mesoscopic semiconductor tunneling heterostructures do%
\cite{esak,tsu}. Such performance may be understood decomposing the system
into coupled molecular subunits that transit collectively through a
conducting quantum resonance as the bias along the device is modified\cite
{han,datta,hettler}. The current-voltage profile is then determined by the
properties of this resonance, which in turn may depend strongly on the
dynamic accumulation of electronic charge in the various subunits, thus
affecting the tunneling rate for transport.

When a pair of neighboring subunits of a different kind are involved, an
energy level associated with one site must reach alignment with one in the
other site in order for a device resonance to be established. The narrow
voltage range in which significant current flows in the experimental data of
Chen et al\cite{chen} suggests that such mechanism may be at work in some
molecules. Using this two site model, Han et al found qualitative agreement
between the model predictions and the experimental data\cite{han}. A time
dependent approach for the analysis of the same model allowed us to fit the
data to an impressive degree of accuracy, and showed in addition that
switching between distinct current states is possible, as reported below.
Switching is extremely fast compared to that in organic complexes, where the
effect has been assigned to local structural disorder induced by the
external electric field\cite{gao}. While in the latter switching times are
of the order of 0.1 $\mu $s, we here report characteristic times in the
order of 0.1 ps, six orders of magnitude shorter. The device is then
sensitive to terahertz pulses, suggesting that the system is potentially
useful as a nanoscopic detector of electromagnetic radiation in the THz
range.

\begin{figure}[t]
\centerline{\includegraphics[width=50mm,angle=-90]{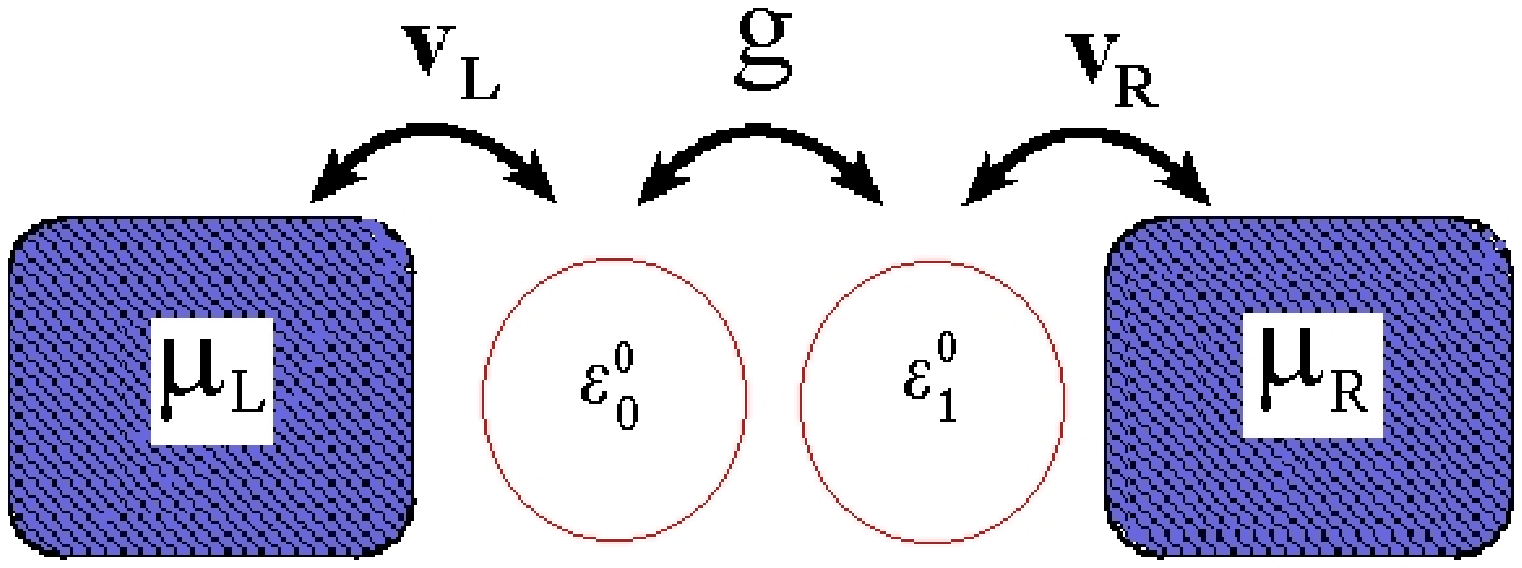}}
\vspace{3mm}
\centerline{\includegraphics[width=50mm,angle=-90]{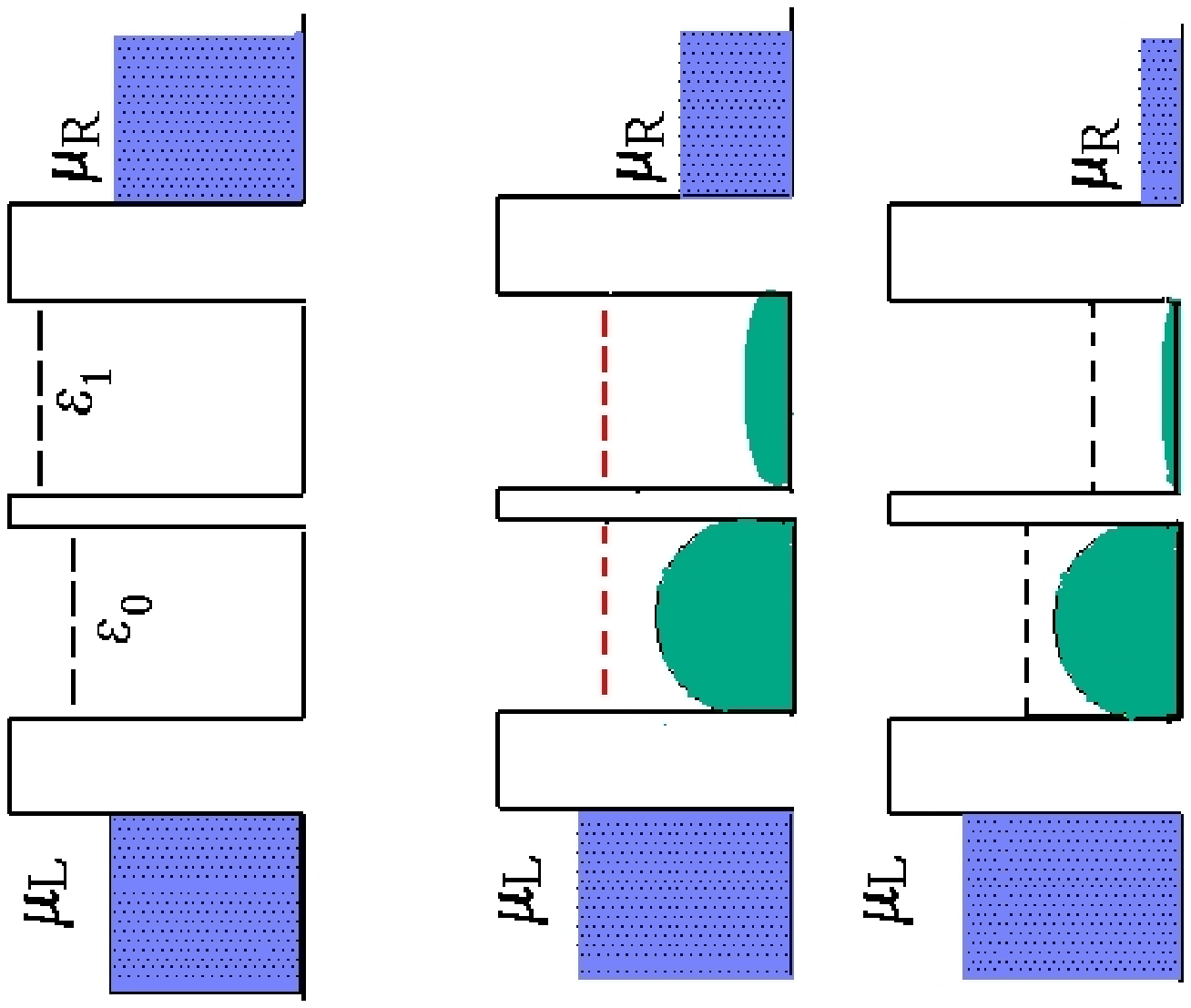}}
\caption{Scheme of the device. a, the molecule is placed between leads at
chemical potentials $\mu _{L}$ and $\mu _{R}$. It is characterized by two
active sites with energies $\varepsilon _{0}^{o}$, $\varepsilon _{1}^{o}$,
which under bias, b, change their relative positions allowing for the
resonance condition to be established in a certain bias range. The shaded
areas show the amount of electron charge at each site. }
\end{figure}

We consider a molecular complex connected to leads subject to an external dc
bias potential. As shown schematically in Fig. 1, the complex is modeled by
two active sites, connected to left and right particle reservoirs at
chemical potentials $\mu _{L}$ and $\mu _{R}$, respectively. The device is
described by the Hamiltonian, 
\begin{widetext}
\begin{eqnarray}
H &=&\sum_{i\neq 0,1;\sigma }\varepsilon _{i}n_{i\sigma }+t\!\!\!\sum_{i\
(\neq 0,1)\ ;\sigma }(\ c_{i,\sigma }^{\dagger }c_{i+1,\sigma
}+c_{i+1,\sigma }c_{i,\sigma }^{\dagger })+\sum_{\alpha =0,1;\sigma
}\varepsilon _{\alpha }n_{\alpha ,\sigma }+v_{L}\sum_{\sigma }(c_{-1,\sigma
}^{\dagger }c_{0,\sigma }+c_{0,\sigma }^{\dagger }c_{-1,\sigma })  \nonumber
\\
&&+v_{R}\sum_{\sigma }(c_{1,\sigma }^{\dagger }c_{2,\sigma }+c_{2,\sigma
}^{\dagger }c_{1,\sigma })+g\sum_{\sigma }(c_{0,\sigma }^{\dagger
}c_{1,\sigma }+c_{1,\sigma }^{\dagger }c_{0,\sigma })+U\sum\limits_{\alpha
=0,1}n_{\alpha _{\uparrow }}n_{\alpha _{\downarrow }}.
\end{eqnarray}
\end{widetext}

\noindent where $c_{i\sigma }^{\dagger }$ ($c_{i\sigma }$ ) creates
(destroys) an electron with spin $\sigma $ at site i and $n_{i,\sigma }=$ $%
c_{i\sigma }^{\dagger }$ $c_{i\sigma }$ represents the occupation of site i.
The molecular sites are labeled 0 and 1 and have intrinsic energies $%
\varepsilon _{0}^{o}$and $\varepsilon _{1}^{o}$, respectively, and an
intra-molecular coupling $g$. The coupling strength of the molecule to the
left (right) lead is $v_{L(R)},$ while t represents the hoping parameter
within the leads. The total site energy $\varepsilon _{i}$ includes the
fixed intrinsic energy at the site, the external radiation-induced voltage
and the applied dc bias, the latter represented by a classical term linear
in the spatial coordinate $i$. Finally, the effect of charge build-up at the
molecular sites is accounted for by a non-linear term with coupling constant
U, the last one in the above expression, that leaves out occupation with the
same quantum numbers. The time-dependent wave function that evolves
according to this Hamiltonian can be expanded in a tight-binding basis as 
\begin{equation}
\left| \psi _{k\sigma }(\tau )\right\rangle =\sum_{i}a_{i\sigma }^{k}(\tau
)\left| \phi _{i\sigma }\right\rangle ,
\end{equation}
\noindent where $\left| \phi _{i\sigma }\right\rangle $ is a Wannier state
of spin $\sigma $ localized at site $i,$ and the coefficients $a_{i\sigma
}^{k}(\tau )$ obey the non-linear equations

\begin{widetext}
\begin{eqnarray}
i\hbar \frac{da_{j,\sigma }^{k}}{d\tau } &=&\varepsilon _{j}(\tau
)a_{j,\sigma }^{k}+t(a_{j-1,\sigma }^{k}+a_{j+1,\sigma }^{k})\;\;\;\;(j\neq
-1,0,1,2),  \nonumber \\
i\hbar \frac{da_{-1(2),\sigma }^{k}}{d\tau } &=&\varepsilon _{-1(2)}(\tau
)a_{-1(2),\sigma }^{k}+v_{L(R)}a_{0(1),\sigma }^{k}+ta_{-2(3),\sigma }^{k}, 
\nonumber \\
i\hbar \frac{da_{0,\sigma }^{k}}{d\tau } &=&(\varepsilon _{0}(\tau
)+U\left\langle n_{0}\right\rangle )a_{0,\sigma }^{k}+v_{L}a_{-1(1),\sigma
}^{k}+ga_{1,\sigma }^{k},  \nonumber \\
i\hbar \frac{da_{1,\sigma }^{k}}{d\tau } &=&(\varepsilon _{1}(\tau
)+U\left\langle n_{1}\right\rangle )a_{1,\sigma }^{k}+v_{R}a_{2,\sigma
}^{k}+ga_{0,\sigma }^{k}.
\end{eqnarray}
\end{widetext}

\noindent In this expression the average site densities equal $\left\langle
n_{0}\right\rangle =\sum_{k,\sigma }\left| a_{0,\sigma }^{k}\right|
^{2},\left\langle n_{1}\right\rangle =\sum_{k,\sigma }\left| a_{1,\sigma
}^{k}\right| ^{2}$, where the sum over $k,\sigma $ covers all occupied
electron states. In writing Eqs. (3) we have adopted a Hartree model for the
electron-electron interaction. As we will show in what follows, the terms
proportional to the site densities generated by this interaction, nonlinear
in nature, are of key importance in the behavior of the system.

Solutions are found discretizing equations (3) and using a half-implicit
numerical method, which is second-order accurate and unitary\cite
{mains,orellana}. It is assumed that the coefficients outside the structure
are given at incident wavenumber $k$ by the expressions

\begin{equation}
a_{j,\sigma }^{k}=(I\ e^{ikja}+R_{j\ }e^{-ikja})\ e^{-i\varepsilon \tau
/\hbar }\;\;\;\;\;\;(ja\le -L)  \label{eq6}
\end{equation}

\begin{equation}
a_{j,\sigma }^{k}=T_{j\ }e^{ik^{\prime }ja}e^{-i\varepsilon \tau /\hbar
}\;\;\;\;\;\;(ja\geq L),  \label{eq7}
\end{equation}

\noindent where the device is explicitly defined to lie between sites $-L$
and $L,$ $a$ is the lattice constant, and $k^{\prime }$ $=[2m(\varepsilon
-V)/\hbar ^{2}]^{1/2}$ is the wavenumber of the transmitted particle, with $%
V=\mu _{L}-\mu _{R}$. The incident amplitude $I$ is supposed to be spatially
constant. Also, far from the molecule the amplitude of the reflected and
transmitted waves $R_{j}$ and $T_{j}$ are supposed to be weakly dependent on
site $j$. This permits to restrict the dependence to the linear term, which
is found to be an adequate approximation provided the time step taken to
discretize equations (3) is less than certain limit value that depends upon
the parametrization of the system. In the numerical procedure the Wannier
amplitudes obtained for one bias are used as the starting point for the next
bias step. Once these coefficients are known the current is calculated from 
\cite{orellana},

\begin{equation}
J_{j}=\frac{e}{h}t\sum_{\sigma
}\int_{o}^{k_{f}}Im(a_{j,\sigma}^{k*}a_{j+1,\sigma }^{k}) dk  \label{eq8}
\end{equation}

\begin{figure}[t]
\centerline{\includegraphics[width=50mm,angle=-90]{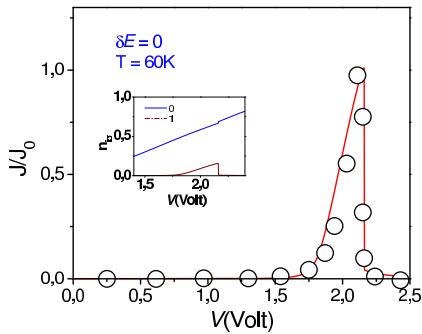}}
\caption{Current-voltage characteristic in the absence of external
radiation. Parameter values are given in the text, and were chosen to fit
the experimental data of Chen et al\protect\cite{chen} at T=60K shown as 
circles. The inset shows the charge in each molecule site at different 
bias values.}
\end{figure}

We apply the above formalism to a specific situation. An excellent fit to
the experimental I-V curve at T=60K of Chen et al.\cite{chen} is obtained in
the absence of radiation within this model, by choosing the parameters $%
\varepsilon _{0}^{o}=0.52,\varepsilon _{1}^{o}=0.92$, $g=0.008,t=2.0,U=0.8$
and intrinsic level broadenings $\Gamma _{L}=\Gamma _{R}=0.16$, all in units
of eV. Here $\Gamma _{L(R)}=\pi v_{L(R)}^{2}\rho (0)$, with $\rho (0)$ the
density of states at the Fermi energy. The energies at the active sites are
represented by $\varepsilon _{i}$ $=\varepsilon _{i}^{o}-[V+\delta E\sin
(2\pi \nu \tau )]/2,$ $($ $i=0,1),$ where $\delta E$ and $\nu $ are the
amplitude and frequency of the incoming THz radiation, respectively.
Energies are measured with respect to the Fermi energy at the emitter
contact ($\mu _{L}$). Figure 2 shows our results, together with experimental
data points from Ref. 5 shown as circles. Notice that the current peak is
quite narrow, and there is a steep fall in the upper critical voltage edge $%
V_{c}$=2.16 V. As the bias is increased and the first site energy approaches
the Fermi energy charge begins to flow, first into site 0, then to site 1 as
well as shown in the inset and, schematically, in Fig. 1b. The effect of
this charge is to raise the local potential at each subunit, thus affecting
the transport coefficient of the overall quantum mechanical structure. The
abrupt fall at $V_{c}$ is caused by a sudden loss of charge at site 1, and
subsequent break of the energy level alignment that allows for resonant
tunneling through the molecule.

\begin{figure}[h]
\centerline{\includegraphics[width=50mm,angle=-90]{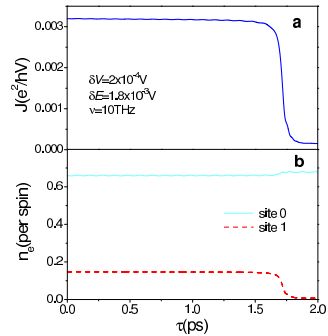}}
\caption{Time evolution of (a) the overall current and (b) the charge
density in sites 0 and 1 at $\nu =10$ THz. Parameters used are as in Fig. 2.}
\end{figure}

Figure 3 shows the time evolution of the current flowing through the system
(Fig. 3a) and the charge density at sites 0 (full line) and 1 (broken line)
(Fig. 3b) when electromagnetic radiation of frequency $\nu =10$ THz and
strength $\delta E=1.8$ mV is turned on at time =0. The offset bias $\delta
V=V_{c}-V$ has the value 0.2mV. Before the radiation enters the sample the
system is in resonance, current flows and both sites have significant
charge. At $\tau $=0 the external field is turned on and after a time of the
order of a picosecond, site 1 looses its charge and the current drops to a
low value. This behavior illustrates the fast switching response of the
device as a THz pulse passes by. The actual transition time is controlled by
the internal coupling constant, so that in our case $T\sim \hbar /g\sim
0.08ps$. There are small oscillations driven by the external field, which we
have found to become larger in amplitude when the signal strength is
increased. The fact that the system undergoes a few coherent oscillations
before falling suggests that the random fluctuations caused by a finite
temperature are not capable of overcoming the inertia of the device. Notice
that in the low current state site 1 has lost its charge. This feature
reveals that the system is bi-stable very close to $V_{c}$. When site 1 is
charged, the local potential is dynamically adjusted to keep the system in
resonance. A disturbance may cause the flow of charge out of the site,
however, with the result that the resonance condition is lost and current
stops.

Suppose now that the system is in a given state of high current, biased
slightly below $V_{c}$. The transition to a low current state is then
possible provided the amplitude of the external field is large enough to
bring the system to criticality. Figure 4 shows the region in which the
transition happens within 2 picoseconds time (labeled YES), separated from
that region in which the radiation is too weak to cause the current to fall
that fast, if at all (labeled NO). The frequency is $\nu =10$ THz and the
bias is shown in terms of the offset from criticality $\delta V=V_{c}-V$. A
rough fit to the boundary is given by $\delta E=0.014\times \delta V^{1/2}$,
all quantities in units of V. Our numerical methods allowed testing
frequencies up to order 1/T, and within that range we found the boundary to
shift to higher values as the frequency increases. In the figure, the filled
circle is a point in the boundary for $\nu =20\ $THz while the triangle is
for $\nu =5$ THz. This is to be expected since in the low frequency limit
one should have $\delta E\sim \delta V.$

\begin{figure}[h]
\centerline{\includegraphics[width=50mm,angle=-90]{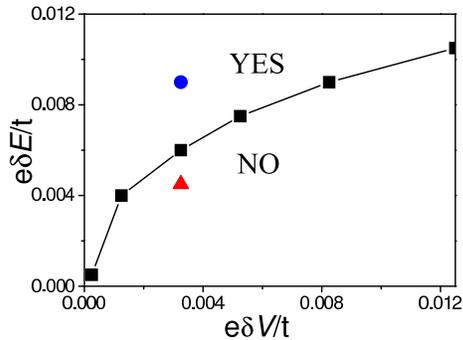}}
\caption{ Boundary separating the region at which switching takes place
within 2 ps of the arrival of an external pulse, labeled YES, from the
region in which the well remains charged and current still flows beyond that
time interval, labeled NO. $\delta V$ measures the external bias relative to
the critical edge $V_{c}$, and $\delta E$ is the radiation strength, both in
units of the lead hopping parameter t. Squares are for incoming radiation of
frequency $\nu =10$ THz. Filled circle and triangle are for frequencies $\nu
=20$ THz and $5$ THz, respectively.}
\end{figure}

In summary, we have shown that switching between different current states in
transport through molecules is possible and may be triggered by a passing
electromagnetic pulse with characteristic times in the terahertz range.
Depending on the applied external bias, the passage of current is turned off
by the radiation pulse, thus allowing the device to act as a tunable
detector of terahertz radiation. Switching occurs because of the existence
of bi-stable states near the I-V characteristic high-voltage edge. A similar
bi-stable region is known to occur in transport through GaAlAs double
barrier resonant tunneling devices\cite{orellana,gold,mart,inko}. A single
active energy level localized in the quantum well formed between the
barriers is responsible for the bi-stability in such case. Near the bottom
of the emitter conduction band a current carrying state through this level,
lifted in energy by the charge trapped in the well, may coexist with a state
of no current at the same bias with the well uncharged. In a two-site
molecular device as the one treated here the effect derives from the dynamic
coupling between the sites within the molecule, and thus has a different
physical origin.

We thank Dr. J.E. Han for valuable discussions and bringing the work of Chen
et al to our attention, and Dr. M. Orszag for comments on the manuscript.
This work was supported in part by FONDECYT grants No. 1020269 and No.
1020829, Programa Milenio ICM P 99-135-F(PO) and C\'{a}tedra Presidencial en
Ciencias (FC).

\end{document}